\documentclass[reprint, amsmath,amssymb, preprintnumbers,aps,nofootinbib,superscriptaddress]{revtex4-1}
\usepackage{graphicx}				
\usepackage{amssymb}
\usepackage{amsmath, amsthm, amssymb}
\usepackage{dcolumn}
\usepackage{bm}
\usepackage{slashed}
\usepackage{tikz}
\usepackage{rotating}
\usepackage{hyperref}
\usepackage{array}
\usepackage{color}
\usepackage{multirow}
\usepackage{xcolor}
\usepackage[T1]{fontenc}
\usepackage{tocloft}
\usepackage[normalem]{ulem}
\usepackage{subcaption}
\usepackage{enumitem}
\usepackage{extarrows}
\usepackage{mathtools}
\usepackage{lipsum}
\usepackage{bbold}
\usepackage{mathrsfs}
\usepackage{lmodern}
\usepackage{blindtext}
\usepackage{lipsum}

\newcommand{\be}{\begin{equation}}
\newcommand{\ee}{\end{equation}}
\newcommand{\bea}{\begin{eqnarray}}
\newcommand{\eea}{\end{eqnarray}}

\begin{document}

\title{PeV Sterile Masses from D-Brane Instantons --- Motivating Stringy Neutrino Options?}

\author{Jim Talbert} 
\affiliation{Niels Bohr International Academy, Niels Bohr Institute, University of Copenhagen, Blegdamsvej 17, DK-2100 Copenhagen, Denmark}

\begin{abstract}
We show that when sterile Majorana neutrino masses are generated non-perturbatively through instantonic interactions in certain classes of string compactifications (e.g. Type IIA orientifold models with intersecting D-branes), then the eigenvalue spectrum for the tree-level mass term $M_p$ can be within the preferred range of the Neutrino Option resolution to the electroweak (EW) hierarchy and neutrino mass problems of the Standard Model, i.e. $M_p \sim$ PeV.  This mechanism holds without tuning for a broad range of string scales, thereby motivating a novel class of string-completed Neutrino Options spanning the light neutrino mass, EW, and deep ultraviolet scales.
\end{abstract}
\maketitle

\section{Introduction and Motivation}
\label{sec:INTRO}
\begin{figure}
\includegraphics[width=85mm]{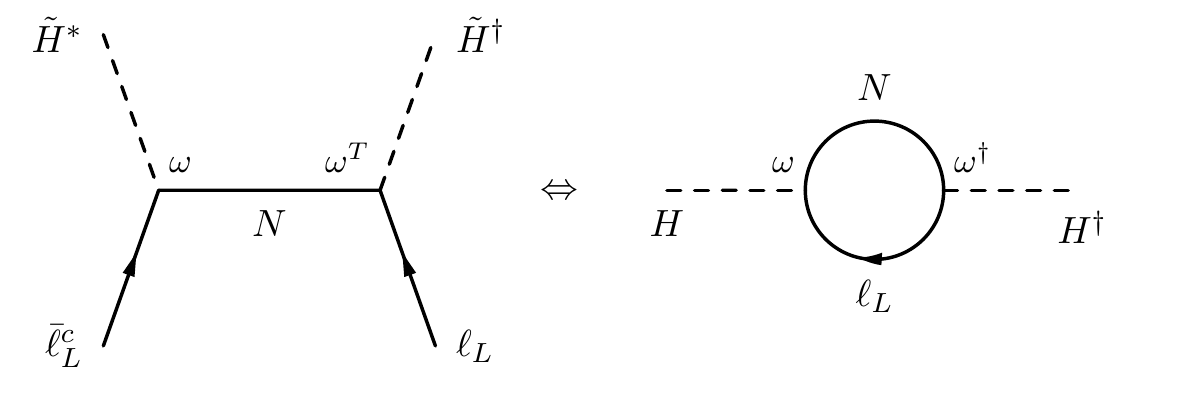}
\caption{Threshold corrections generating the EW and neutrino mass scales in the Neutrino Option. 
}
\label{fig:match}
\end{figure}

The Neutrino Option \cite{Brivio:2017dfq,Brivio:2018rzm} is a paradigm for simultaneously resolving the electroweak (EW) hierarchy and neutrino mass problems of the Standard Model (SM) by assuming that the dominant threshold corrections to the mass of the Higgs scalar, and therefore also the generation of the EW scale, come from integrating out heavy Majorana masses $M_p$ in the effective field theory (EFT) defined by the seesaw mechanism for light neutrino mass generation \cite{Minkowski:1977sc,GellMann:1980vs,Mohapatra:1979ia,Yanagida:1980xy}.  Indeed, it is inevitable that when a Type-I seesaw Lagrangian of the form 
\begin{equation}
\label{eq:LagSS}
\mathcal{L}_N = \frac{1}{2} \overline{N}_p \left(i \slashed{\partial} - M_p \right) N_p - \left[ \overline{l^\beta_L} \tilde{H} \omega_\beta^{p,\dagger} N_p + \overline{N}_p \omega_\beta^p \tilde{H}^\dagger l^\beta_L \right] \,,
\end{equation}
with $N_p = N_p^c$ the mass-eigenstate sterile Majorana state defined by  \cite{Bilenky:1980cx,Broncano:2002rw}
\begin{equation}
N_{p} = e^{i \theta_p /2} N_{R,p} + e^{-i \theta_p /2} (N_{R,p})^c \,,
\end{equation}
is matched to the SMEFT after integrating out heavy $N_p$,
a threshold correction to the Higgs boson mass of the form
\begin{equation}
\label{eq:Higgsthresh}
-\frac{m_0^2}{2} \, H^\dagger H \longrightarrow \left(-\frac{m_0^2}{2}  - \frac{M_p^2 \vert \omega_p \vert^2}{16 \pi^2} \right) H^\dagger H
\end{equation}
necessarily appears at one-loop order --- see Figure \ref{fig:match} for Feynman diagrams relevant to this matching, and \cite{Elgaard-Clausen:2017xkq} for a complete tree-level analysis up to dimension seven.  Here $m_0$ is the bare Higgs mass, $\theta_p$ is an arbitrary phase, and the $p,\beta \in \lbrace 1,2,3 \rbrace$ indices in \eqref{eq:LagSS}-\eqref{eq:Higgsthresh} are respectively Majorana and SM lepton flavour indices.  The Higgs threshold correction in \eqref{eq:Higgsthresh} of course occurs in addition to the seesaw generation of the active light neutrino masses $m_\nu$, such that 
\begin{equation}
m_\nu \sim \frac{\omega_p^2 \bar{v}_T^2}{M_p}, \,\,\,\,\,\, m_h \sim \frac{\omega_p M_p}{4 \pi}, \,\,\,\,\,\, \bar{v}_T \sim \frac{\omega_p M_p}{4 \sqrt{2} \pi \sqrt{\lambda}},
\end{equation}
where $\sqrt{2 H^\dagger H} \equiv \bar{v}_T$ defines the EW scale and $\lambda$ is the Higgs self coupling.  As noted by Vissani \cite{Vissani:1997ys}, the correction to $m_h$ can represent a manifestation of the hierarchy problem in the context of the seesaw model, if extreme fine-tuning is required to realize the experimental value of $m_h \sim 125$ GeV.  However, in the Neutrino Option (with degenerate or only mildly hierarchical $M_p$) one instead recognizes that with the parameter space \cite{Brivio:2017dfq,Brivio:2018rzm},\footnote{This estimate was made in the simplifying limit of $M_1 = 0$, which yields a successful seesaw, but can also be generalized to a complete three-eigenvalue analysis, with the expectation that (assuming minimal hierarchies in $\omega_\beta^p$) $M_1 \lesssim M_{2,3}$ as well, in order to avoid unnatural threshold corrections to $m_h$. The bounds are also largely stable against variations of other input parameters, e.g. the top quark mass and neutrino mixing elements.}
\begin{align}
\label{eq:parameterspace}
M \lesssim 10^{4} \,\,\, \text{TeV} = 10 \,\,\, \text{PeV}, \,\,\,\,\, \text{with} \,\,\,\,\, \vert \omega \vert \simeq \frac{\text{TeV}}{M} \,,
\end{align}
the threshold corrections associated to $M_p$ can instead \emph{minimize} fine-tuning by in fact defining and generating the observed  $\bar{v}_T$.  In this way the same mechanism explains the origins of both the EW and neutrino mass scales, and in a manner which can also accommodate successful resonant leptogenesis if $M \gtrsim 1$ PeV \cite{Brivio:2019hrj} (also see \cite{Brdar:2019iem}). Hence an important theoretical question to address is how the $M_p \sim$ PeV scale can appear naturally, perhaps from the deep ultraviolet (UV) typically associated to Grand Unified (GUT) or Planckian dynamics.

In \cite{BTTpert} the loop quantum corrections to $M_p$ permitted by \eqref{eq:LagSS} were studied, with the conclusion that the Neutrino Option is perturbatively stable; neither threshold corrections nor renormalization group flow can generate the PeV scale from a heavier Majorana state with (e.g.) $M_3 \gg$ PeV in the context of minimal Type-I seesaw models, and indeed in many Beyond-the-Standard Model (BSM) extensions to the Type-I seesaw.  Permitting extra scalar fields, the mass term could appear as a result of spontaneously broken classical scale invariance, leading to a conformal Neutrino Option as in \cite{Brdar:2018vjq}.  On the other hand, non-perturbative dynamics might also provide a route to a successful, minimal UV Neutrino Option, and in this letter we study a particular instantiation of this non-perturbative option by focusing on a well-known mechanism \cite{Ibanez:2006da,Ibanez:2007rs,Blumenhagen:2006xt,Cvetic:2007ku} for generating sterile Majorana neutrino masses $M_p$ from instanton interactions in a large class of string compactifications.  We will show that such stringy instantons can in fact naturally generate the preferred $M_p \sim$ PeV scale given a broad range of string scales, and therefore a potentially broad range of string embeddings.
\section{Stringy Majorana Mass Matrices}
\label{sec:MASS}
In a series of papers \cite{Ibanez:2006da,Ibanez:2007rs} (also see \cite{Blumenhagen:2006xt,Cvetic:2007ku}) it was shown that a Majorana mass term of the form
\begin{equation}
\label{eq:massgen}
 m_s\, e^{-U}\,\overline{N} N
\end{equation}
can appear as a result of instanton effects in string compactifications to four-dimensional (4D) SM and minimally-supersymmetric SM (MSSM) spectra. Here $U$ is a set of complex string moduli and $m_s$ is the string scale, and the term is allowed under the SM gauge symmetry and, as is ubiquitous in many classes of string compactifications, an additional U(1)$_{B-L}$ gauge symmetry.  This latter invariance is due to the fact that $\text{Im}\,U$ is (a linear combination of) axion-like complex scalar fields which is also shifted by two units under $B-L$, thereby leaving the term invariant.  The $U(1)_{B-L}$ gauge boson obtains a mass on the order of $m_s$ through a Stueckelberg mechanism, leaving in the `infrared' (IR) only an unbroken global $U(1)_{B-L}$ and the gauge symmetry and light particle spectrum of the SM (or e.g. the MSSM) \cite{Ghilencea:2002da}.\footnote{Hence $M_{Z^\prime} \sim m_s$ is a disconnected scale unrelated to a vacuum expectation value \cite{Heeck:2014zfa}, and 
any kinetic mixing between $U(1)_{B-L}$ and $U(1)_Y$ sectors typical of Stueckelberg-extended models (see e.g. \cite{Kors:2004dx,Kors:2004iz,Perez:2014gta,Feldman:2011ms,Salvioni:2009mt}) is suppressed by $m_s$, and may not even be present in the relevant classes of D-brane models \cite{Ibanez:2001nd,Kors:2004iz}. Of course, if in a given model the $Z^\prime$ couples to $B$, two-loop Higgs corrections can put naturalness constraints on $m_s$ or the unknown couplings.}  In what follows we assume that no 4D SUSY is present, although this implies important constraints on the string embedding --- see the discussion in Section \ref{sec:SUSYcomment}.

For example, it was shown in \cite{Ibanez:2006da} that \eqref{eq:massgen} can appear in Type IIA orientifolds where fermions appear as localized string excitations at the intersection of D6-branes, and where two such D6-branes further intersect with D2-branes to be identified with the string instanton, giving rise to fermionic zero modes (Grassman variables) that must be integrated over.  This integration yields a non-vanishing contribution to the fermion ($N$) bilinear amplitude which, upon multiplying by the classical D2 instanton action, yields a Majorana mass matrix given by\footnote{Similar instanton effects can also generate a Weinberg operator whose relative strength with respect to \eqref{eq:massgenB} is of phenomenological relevance, but is also model-dependent.  In what follows we assume that \eqref{eq:massgenB} is the dominant contribution to the light neutrino spectrum.}

\begin{equation}
\label{eq:massgenB}
M_{ab} = m_s \left(\epsilon_{ij} \epsilon_{kl} d_a^{ik} d_b^{jl} \right) \, e^{-S}\,,
\end{equation}
where $d$ represent flavoured couplings to be discussed below, $\epsilon_{ij}$ is the antisymmetric unit tensor, and $S$ is  the instanton action with suppression factor $\text{Re}(S)$. However, the D2-instantons giving rise to this also exhibit a global internal symmetry in 4D, and in \cite{Ibanez:2007rs} a somewhat exhaustive scan of available instantons/symmetries was performed.\footnote{While \cite{Ibanez:2007rs} was largely looking for MSSM-like spectra, the appearance of the internal $Sp(2)$ is an independent phenomenon and may very well appear in similar scans with no 4D SUSY.  I thank Luis Ib\'a\~nez for this clarifying comment.}
The class that appeared most frequently had an internal $Sp(2) \cong SU(2)$ symmetry that leads to a form of \eqref{eq:massgenB} whose flavour structure factorizes according to 
\begin{equation}
\label{eq:MRa}
M_{ab} = 2\, m_s \sum_r d^{(r)}_a d^{(r)}_b e^{-S_r}\,,
\end{equation}
such that
\begin{widetext}
\begin{equation}
\label{eq:MRdem}
M = 2 m_s \sum_r e^{-S_r} \text{diag} \left(d_1^{(r)}, d_2^{(r)}, d_3^{(r)} \right) \cdot \left(
\begin{array}{ccc}
1 & 1 & 1\\
1 & 1 & 1\\
1 & 1 & 1
\end{array}
\right) \cdot \text{diag} \left(d_1^{(r)}, d_2^{(r)}, d_3^{(r)} \right)\,.
\end{equation}
\end{widetext}
The sum over $r$ accounts for the fact that, in general, multiple instantons can contribute to the bilinear amplitude, and it is clear that when $r$ instantons are present, there are $3-r$ zero eigenvalues in \eqref{eq:MRdem}.  Note however that \eqref{eq:massgenB}-\eqref{eq:MRdem} are not in their mass-eigenstate basis, and so $r \neq p$.  Furthermore, \eqref{eq:MRdem} has three inputs associated to it:  $m_s$, $S_r$, and $d_a^{(r)}$.  Following the discussion in \cite{Ibanez:2006da,Antusch:2007jd}, they constitute:
\begin{itemize}
\item ${\bf{m_s}}$:  The \emph{string scale} is not known, and could in principle be as low as current LHC bounds allow, although in the supersymmetric case one might want to keep this scale high (e.g. the GUT scale) to maintain coupling unification, etc.  Regardless, 4D SUSY is not required for the mechanism to work (see \cite{Ibanez:2006da}) and we generally assume its absence here,  so when the seesaw mechanism is the dominant source of IR neutrino masses (as opposed to the Weinberg operator) as desired in the Neutrino Option, $m_s$ is essentially unconstrained.
\item ${\bf{S_r}}$: The \emph{instanton action} is, critically, not as large a suppression factor as in standard Yang Mills (e.g. electroweak) instanton effects, which typically go $\propto e^{-1/g_2^2}$.  In fact, there are no phenomenological constraints on this parameter, such that the overall suppression could be as small as an $\mathcal{O}(1)$ effect.  For the phenomenology of the next section to work, we will find that $ \vert \text{Re}(S_r) \vert \sim \mathcal{O}(10^{-1}-10^{1})$ is desirable, although this depends on the string scale.  Since the overall normalization of the action is a free parameter in (at least some classes of) perturbative string compactifications, this seems quite safe, and is in fact consistent with the neutrino phenomenology that was studied in \cite{Antusch:2007jd}.
\item ${\bf{d_a^{(r)}}}$:   The \emph{flavour vectors} are fundamentally stringy objects, and in certain classes of compactifications they are computable, as are normal Yukawa couplings (see e.g. \cite{Cremades:2003qj}).  However, they are also model-dependent, so the authors of \cite{Antusch:2007jd} took a phenomenological perspective and allowed these to be constrained by data.  As we want to be as model-independent as possible, we will do the same below.
\end{itemize}
Of course, as mentioned in \cite{Ibanez:2006da}, this stringy instanton mechanism is quite general and may lead to Majorana mass matrices of the form in \eqref{eq:massgen} in multiple classes of string compactifications (not just the Type IIA models referenced above). Hence for our phenomenological purposes we simply take \eqref{eq:massgen}-\eqref{eq:massgenB}, with the particular factorization in \eqref{eq:MRa}, as the starting point of our analysis. However, we emphasize that one does not require the special democratic factorization in \eqref{eq:MRa} for the Neutrino Option to be successful --- any sterile mass matrix with $M_p \sim$ PeV, including a rank-two matrix with $M_1 = 0$, can work.  However, \eqref{eq:MRa} allows an intuitive understanding of the role of each instanton in the tree-level non-perturbative mass generation, and further allows for an easy qualitative extraction of relevant UV scalings of $m_s$, $d_a^{(r)}$, and $S_r$ from IR neutrino phenomenology in Section \ref{sec:PHENO}.
\subsection{A Comment on Finite String Embeddings}
\label{sec:SUSYcomment}
Before proceeding though, an important comment/caveat must be mentioned in order for the framework to be fully consistent.  As is obvious from \eqref{eq:Higgsthresh}, a successful Neutrino Option relies on the fact that the threshold corrections to $m_h$ from $N_p$ are in fact dominant (or at least not sub-dominant), such that the origin of the EW scale is both explained and remains stable.  However, in non-SUSY string embeddings one might expect potentially destabilizing radiative corrections to $m_h$ on the order of $\alpha_{SM}\, m_s^2$, driven by Kaluza-Klein or string excitations of SM states.  Furthermore, solutions to moduli stabilization which result in a 4D EFT with a massive SUSY spectrum could circumvent the Neutrino Option altogether --- either the sparticles are roughly TeV scale and there is no hierarchy problem (assuming safe radiative corrections from sneutrino partners to $N_p$), or they are extremely heavy and can instead \emph{generate} the hierarchy problem, with or without heavy $N_p$ (although see \cite{Giudice:2011cg} for scenarios where this is not necessarily the case, when SUSY is broken at an extremely high scale).  The former option is increasingly challenged by the lack of LHC SUSY signals, and the latter is not solved by the Neutrino Option.   

Hence it is critical that SM excitations are stabilized in the non-SUSY string embedding our present mechanism assumes, or in any other embedding where \eqref{eq:Higgsthresh} is maintained as dominant, such that a non-standard `Neutrino Option' exists.  One option may be to simply assume a low value of the string scale, as was done in \cite{Antoniadis:1998ig}, although the requirement of large extra dimensions in this scenario introduce additional considerations in D-brane models --- see e.g. \cite{Cremades:2002qm} --- and the preferred range of $m_s \in \lbrace 1 - 10 \rbrace$ TeV of \cite{Antoniadis:1998ig} likely requires `non-perturbative' flavour vectors in the Neutrino Option.  Regardless, in Section \ref{sec:NUMERICS} we study low(er)-scale $m_s$ scenarios that may naively be more compatible with non-SUSY constructions, and show that without further theory constraints they are also viable.  Of course other efforts to stabilize moduli in non-SUSY compactifications exist (see e.g. \cite{Blumenhagen:2002mf} which addresses intersecting D-branes in Type 0 string theory, and \cite{Blumenhagen:2006ci} for a review of other ideas), and it was already mentioned in \cite{Ibanez:2006da} that the introduction of suitable fluxes in the compactification (see e.g. \cite{Marchesano:2006ns}) could lift additional zero modes of the D2-Brane instantons, thereby leaving their deformation moduli stable. However, ultimately we are unaware of any mechanism that has demonstrated complete stability in the Type IIA  non-SUSY intersecting D-brane models of \cite{Ibanez:2006da}, or other scenarios that have been shown capable of explicitly generating the SM $+N_p$ spectrum of interest --- whether this can be done remains an open question in the string literature.\footnote{I am grateful to Graham Ross for pointing out the general concern of SM string excitations in the Neutrino Option, and to Angel Uranga for helpful commentary on the present literature.}

In summary, for our phenomenological purposes here, we simply assume a finite string embedding from the outset,
and leave stabilization as a potentially limiting theoretical constraint on successful ultraviolet stringy Neutrino Options.  On the other hand, the compelling observations of the upcoming Sections \ref{sec:PHENO}-\ref{sec:NUMERICS} also provide a novel motivation for further pursuing such embeddings, as we show that, if found, the resulting theory can still address the EW hierarchy and neutrino mass problems within the Neutrino Option paradigm, i.e. in a manner previously unconsidered.
\section{Low Energy Neutrino Phenomenology and Implied Scales}
\label{sec:PHENO}
Gaining complete analytic control over the Type-I seesaw in this framework is challenging.  However, given \eqref{eq:MRa}-\eqref{eq:MRdem}, the low-energy neutrino mass matrix $M^L$ is given, in the simplifying limit of a diagonal neutrino Yukawa coupling,\footnote{Small perturbations about this special case can of course be systematically studied, but the  qualitative range of $M_p$ we are attempting to identify here becomes especially manifest in this diagonal limit.  Note however that \eqref{eq:IRMLmodel}-\eqref{eq:IRMLa} assumes the dominant leptonic mixing originates from $M$, which is not necessary for our analysis, and which we relax in Section \ref{sec:NUMERICS}.} by 
\begin{equation}
\label{eq:IRMLmodel}
M_{ab}^L = \left(\sum_r \frac{d_a^{(r)}}{(h_D^T)_{aa}} \frac{d_b^{(r)}}{(h_D)_{bb}} \tilde{I}_r^{-1} \right)^{-1}\,,
\end{equation}
with 
\begin{equation}
h_D = \text{diag} \left(\omega_e^{(\nu)}, \omega_\mu^{(\nu)}, \omega_\tau^{(\nu)} \right),\,\,\,\,\,\,\,\,\,\,\text{and} \,\,\,\,\,\,\,\,\,\, \tilde{I}_r = \frac{\langle \overline{H} \rangle^2}{2 m_s} \frac{1}{e^{-S_r}}\,.
\end{equation}
However, we also recall that in this same basis one can generically derive that, in the further limit of three non-zero masses, 
\begin{equation}
\label{eq:IRMLa}
\left(M_{ab}^L\right)^{-1} = U \cdot \text{diag} \left(\frac{1}{m_1\, e^{i \alpha_1}}, \frac{1}{m_2 \, e^{i \alpha_2}}, \frac{1}{m_3}\right) \cdot U^T \,,
\end{equation}
since $M_{ab}^L$ is a complex symmetric matrix.  Assuming diagonal charged-lepton mixing (which is always possible with an appropriate basis transformation), $U$ is then the PMNS matrix, up to the diagonal matrix of Majorana phases $\alpha_i/2$ that has simply been factored into the IR neutrino mass eigenvalues $m_i$.  We now see that the RHS \eqref{eq:IRMLa} can be expanded as 
\begin{equation}
\left(M_{ab}^L\right)^{-1} =  \frac{e^{-i \alpha_1}}{m_1}\, U_{i1}\cdot U_{1i} +  \frac{e^{-i \alpha_2}}{m_2} \,  U_{i2}\cdot U_{2i} +  \frac{1}{m_3} \, U_{i3}\cdot U_{3i} \,.
\end{equation}
Equating this form of the IR mass matrix to \eqref{eq:IRMLmodel}, one immediately arrives at a solution for $\tilde{I}_r$ in terms of the low energy neutrino mass eigenvalues given by
\begin{equation}
\label{eq:massequate}
\tilde{I}_1 = m_1\,, \,\,\,\,\,\,\,\,\,\,\tilde{I}_2 = m_2\,, \,\,\,\,\,\,\,\,\,\, \tilde{I}_3 = m_3\,,
\end{equation}
and a solution for the components of the flavour vectors $d_a^{(r)}$ in terms of the PMNS mixing matrix given by
\begin{align}
\nonumber
\left(\frac{d_1^{(1)}}{\omega_e^{(\nu)}}, \frac{d_2^{(1)}}{\omega_\mu^{(\nu)}}, \frac{d_3^{(1)}}{\omega_\tau^{(\nu)}} \right) &= e^{-i \alpha_1/2}\left(U_{11}, U_{21}, U_{31} \right)\,,\\
\nonumber
\left(\frac{d_1^{(2)}}{\omega_e^{(\nu)}}, \frac{d_2^{(2)}}{\omega_\mu^{(\nu)}}, \frac{d_3^{(2)}}{\omega_\tau^{(\nu)}} \right) &= e^{-i \alpha_2/2}\left(U_{12}, U_{22}, U_{32} \right)\,\\
\label{eq:PMNSequate}
\left(\frac{d_1^{(3)}}{\omega_e^{(\nu)}}, \frac{d_2^{(3)}}{\omega_\mu^{(\nu)}}, \frac{d_3^{(3)}}{\omega_\tau^{(\nu)}} \right) &= 
\left(U_{13}, U_{23}, U_{33} \right)\,.
\end{align}
In \eqref{eq:IRMLmodel}-\eqref{eq:PMNSequate} we have simply generalized the mixing matrix elements from the specific texture originally studied in \cite{Antusch:2007jd}.  As $U_{PMNS}$ is unitary, the matrix elements are all  $\vert U_{ij} \vert \le 1 $ and the Majorana phase factor $e^{-i \alpha_i/2}$ is also $\mathcal{O}(1)$.  Hence we derive that
\begin{equation}
\label{eq:djScale}
d_a^{(r)}  \approx  \omega_i^{(\nu)}/10 \sim \mathcal{O}(10^{-5}) - \mathcal{O}(10^{-4}),
\end{equation}
where we generically assume non-hierarchical couplings, which is consistent with the original Neutrino Option analysis in \cite{Brivio:2017dfq,Brivio:2018rzm}  and is also more preferable from the stringy side \cite{Antusch:2007jd}, and where the scale relationship on the RHS is the range identified in \eqref{eq:parameterspace}.

Let us now derive the expected scale of the remaining parameter that enters \eqref{eq:MRa}-\eqref{eq:MRdem}, the instanton suppression factor $S_r$, given (e.g.) a string scale $m_s$ assumed to lie in the deep UV.  Using \eqref{eq:massequate}, one easily finds that  
\begin{equation}
\label{eq:Sr&mr}
S_r = -\frac{1}{2} \ln \left[ \frac{\langle \overline{H} \rangle^4}{m_r^2 \, 4 \, m_s^2} \right]\,\,\, \Longleftrightarrow \,\,\, m_r^2 = \frac{\langle \overline{H} \rangle^4}{4 m_s^2}\,e^{2 S_r}\,,
\end{equation}
from which we also obtain the neutrino mass-squared difference formula
\begin{equation}
\label{eq:massdifmodel}
\Delta m_{r_i r_j}^2 \equiv m_{r_i}^2 - m_{r_j}^2 = \frac{\langle \overline{H} \rangle^4}{4 m_s^2}\,\left(e^{2 S_i} - e^{2 S_j}\right)\,.
\end{equation}
This quantity is of course experimentally bound by neutrino oscillation experiments.  The most recent global fit from the {\tt{NuFIT}} collaboration \cite{Esteban:2020cvm} puts these at 
\begin{align}
\nonumber
\text{(NO):}\ \ \ \ \ \ \Delta m_{21}^2 &\in \lbrace 6.82 - 8.04 \rbrace \,\,\,\cdot \,\,\,\, 10^{-5} \,\,\,\,\, \text{[eV]}^2 \,, \\
\ \ \ \ \ \ \,\,\,\,\,
\label{eq:NuFITmass}
 \Delta m_{31}^2 &\in \lbrace 2.435 - 2.598  \rbrace \,\cdot 10^{-3} \,\,\,\,\, \text{[eV]}^2\,,
\end{align}
where we have shown the $3 \sigma$ range of values when the neutrino masses are of normal ordering (NO).  Inverted ordering (IO) fits are also available, as are global bounds on the PMNS elements $U_{ij}$ in \eqref{eq:PMNSequate}, although as mentioned above these will all be $\mathcal{O}(U_{ij}) \lesssim 1$ due to unitarity in our theoretical framework.  So for the order of magnitude estimate we attempt here we will not repeat the explicit bounds.
Hence, using a deep-UV $m_s$,
\begin{align}
\label{eq:paramset}
m_s \sim \lbrace 10^{16} - 10^{19} \rbrace \,\,\,\,\, \text{GeV}\,, \,\,\,\,\,\langle \overline{H} \rangle \sim 246 / \sqrt{2} \,\, \,\,\, \text{GeV} \,,
\end{align}
we can then use \eqref{eq:Sr&mr}-\eqref{eq:NuFITmass} to derive the rough order of magnitude estimate
\begin{equation}
\label{eq:SrScale}
\vert S_r \vert \sim \mathcal{O}(10^{-1}-10^1)\,.
\end{equation}
This provides all of the information necessary to then determine the expected scale of $M$.

To that end we can insert \eqref{eq:djScale}, \eqref{eq:paramset}, and \eqref{eq:SrScale} into \eqref{eq:MRa}-\eqref{eq:MRdem}, and compute the eigenvalue spectrum of $M$. As a first illustration we allow for three instantons and choose the parameters randomly within the ranges specified above.  For example, with $m_s = 10^{18}$ GeV one can find 
\begin{align}
\nonumber
M_3 &\sim \mathcal{O}(10^6-10^{10}) \,\,\,\,\,\, \text{GeV},\\
\nonumber
M_2 &\sim \mathcal{O}(10^4-10^9) \,\,\,\,\,\,\,\, \text{GeV},\\
\label{eq:PeV}
M_1 &\sim \mathcal{O}(10^2-10^8) \,\,\,\,\,\,\,\, \text{GeV}\,.
\end{align}
While this is only a qualitative estimate, made within the simplified formalism implied by \eqref{eq:IRMLmodel}, the obvious implication is that the PeV scales preferred in the Neutrino Option can appear automatically in this formalism, given GUT or Planckian string scales $m_s$ and values implied by low energy neutrino phenomenology.  Hence \eqref{eq:PeV} and its analogues across a broader range of $m_s$ represent the core observation of this note, which we now study in more numerical detail. 

\section{Scanning Viable Parameter Spaces}
\label{sec:NUMERICS}
\begin{figure}[t]
\includegraphics[width=90mm]{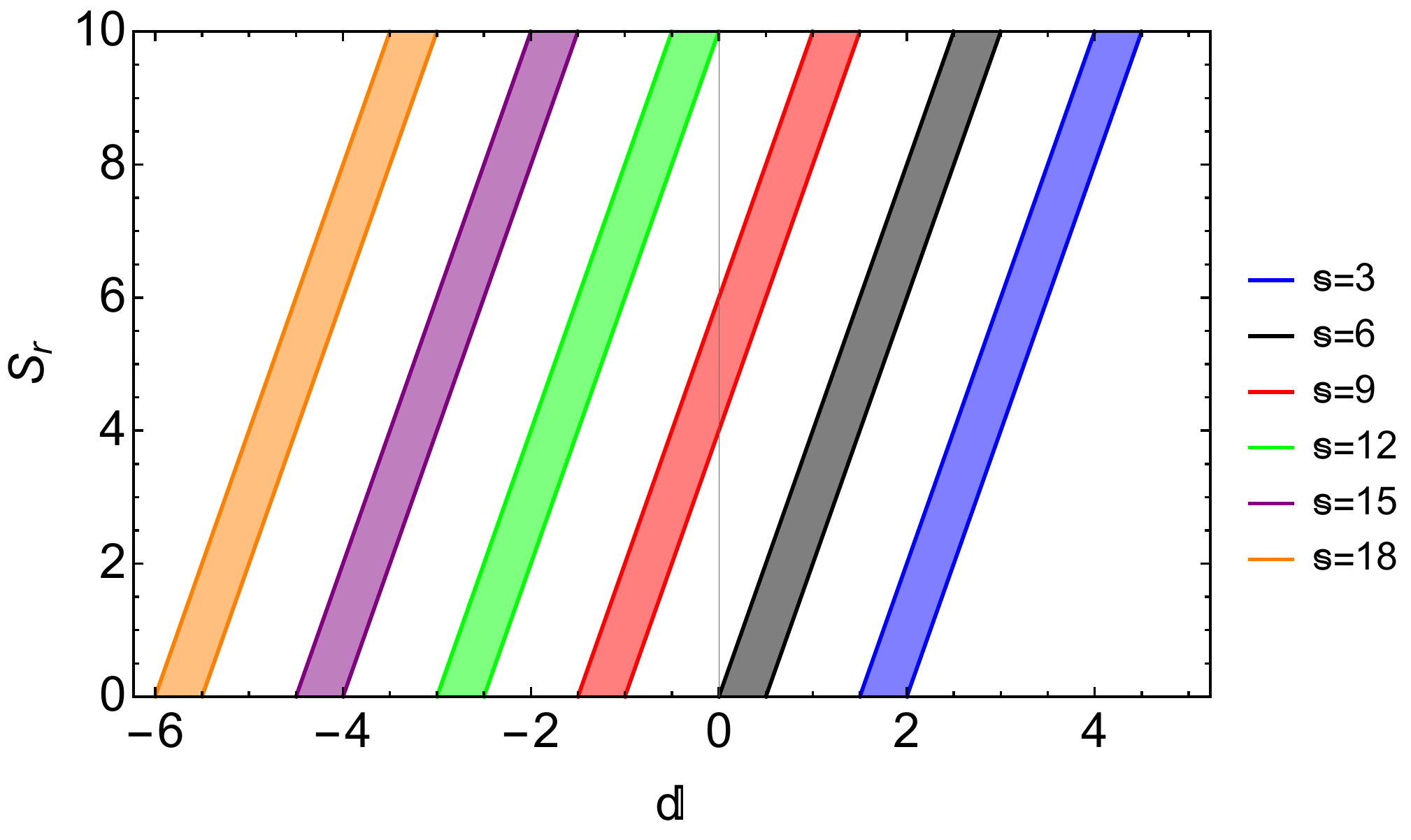}
\caption{Contours realizing (roughly) successful Neutrino Options across a broad range of string scales, given the scalings of the free parameters considered in the text.}
\label{fig:contourscale}
\end{figure}
We now explore the viable parameter space more comprehensively by varying the parameters relevant to calculating the  $M_p$ spectrum, which constitute the string scale $m_s$, the instanton suppression factors $S_r$, and the flavour vectors $d^{(r)}_a$.
Since our goal is to show that successful Neutrino Options can be realized with a broad range of string scales, we define the following scale-setting parameters
\begin{equation}
\label{eq:logstring}
\mathbb{s} \equiv \log_{10} m_s, \,\,\, \mathbb{d} \equiv \log_{10} d_a^{(r)}, \,\,\, \mathbb{p}  \equiv \log_{10} M_p,
\end{equation}
with dimensionful quantities given in GeV, and note that, up to $\mathcal{O}(1)$ coefficients and flavour effects, naive dimensional analysis implies the following contour:
\begin{equation}
\label{eq:gencontour}
\mathbb{s} + 2 \mathbb{d} - \left(\frac{1}{2} S_{r}  + \mathbb{p}\right) \approx 0 \,,
\end{equation}
where we assume non-hierarchical flavour vectors, but do \emph{not} assume the strict mixing limits of Section \ref{sec:PHENO}.\footnote{Here we have slightly abused notation, setting $\text{Re}(S_r) = S_r$.  In the upcoming numerical analysis, we will treat $S_r$ as a complex number with randomly varied real and imaginary components.}  A successful Neutrino Option then fixes $\mathbb{p}$ as in \eqref{eq:parameterspace}, such that \eqref{eq:gencontour} defines a hyperplane with coordinates $\mathbb{s}$, $\mathbb{d}$, and $S_r$.  For the instanton suppression factors, one notices from the mass-difference formula of \eqref{eq:massdifmodel} (and more generally from \eqref{eq:MRdem}) that $r > 1$ is required to reproduce low-energy neutrino data.  This can be achieved naturally since each $S_r$ is associated to a different instanton, as can hierarchies amongst the individual $S_r$, facts that were first noted in \cite{Antusch:2007jd}.  In what follows we consider the $r=3$ case, such that there are three non-zero eigenvalues in $M$. 
Continuing, the string parameter $\mathbb{s}$ is only bound by experiment --- low values are interesting in general, but especially so due to the potential stability issues mentioned in Section \ref{sec:SUSYcomment} for non-SUSY string embeddings, and of course high values provide a direct link to the deep UV without relying on large compact dimensions.  
Finally, we remain agnostic as to the possible values of $\mathbb{d}$ (which are extremely model-dependent),  and allow for the following broad range of scalings:
\begin{equation}
\label{eq:drange}
-6 \le \mathbb{d} \le 5\,.
\end{equation}
The lower bound is motivated by Figure \ref{fig:contourscale}, \eqref{eq:djScale}, and the simultaneous fact that $d^{(r)}_a$ are computed analogously to standard Yukawa couplings in certain types of toroidal compactifications, including those of \cite{Ibanez:2006da}.  Assuming $\text{Re}(S_r) > 0$ (such that one truly has an instanton \emph{suppression} factor), one immediately recognizes from \eqref{eq:parameterspace} that the models of \cite{Antoniadis:1998ig}, with $\mathbb{s} \in \lbrace 3, 4 \rbrace$, are not realized in this limit.  On the other hand, for $\mathbb{d} \ge 0$, low-scale models are viable.  While one might suspect that $d^{(r)}_a \lesssim1$ in order to remain perturbative (motivated by the analogy with Yukawa couplings), we recall that \eqref{eq:massgenB} is a non-perturbative operator, and may not necessarily be constrained by such arguments.

Given these inputs, we plot the naive scaling contours expected for viable Neutrino Options in Figure \ref{fig:contourscale}, for string scales ranging from 1 TeV to 10$^{18}$ GeV, and where we have treated $S_r$ in \eqref{eq:gencontour} as a single $\mathcal{O}(1)$-$\mathcal{O}(10)$ parameter, which is varied over a significant (positive) range to account for \eqref{eq:Srelate}.  The width of the colored bands represent the bounds on $M_p$ from \eqref{eq:parameterspace}, and allowing for continuous string scales between the discretized values shown, one notes that, without further theory constraints from a specific string embedding/model, a broad range of parameter values can yield successful Neutrino Options in the framework.  

To better illustrate this, we now account for the various $\mathcal{O}(1)$ coefficients and flavour structures in \eqref{eq:MRdem} by defining the components of the flavour vectors as
\begin{equation}
\label{eq:dparam}
d_{a}^{(r)} \equiv \left(\rho \,e^{i \phi} \,\cdot \,10^{\mathbb{d}_0}\right)_a^{(r)} \,,
\end{equation}
such that $\mathbb{d}_0 \sim \mathbb{d}$, and $\lbrace \rho,\phi \rbrace $ simply parameterize the arbitrary $\mathcal{O}(1)$ complex coefficient.  Then, in order to avoid tuning amongst the $S_r$ and avoid large hierarchies in $M_p$ (in accordance with the analysis in \cite{Brivio:2017dfq,Brivio:2018rzm}), we impose the following simple scaling relationship,
\begin{equation}
\label{eq:Srelate}
S_{2,3} \overset{!}{=} c_{2,3} \, S_1, \,\,\,\,\, \text{with} \,\,\,\,\, c_{2,3} \sim \mathcal{O}(1),
\end{equation}
where $c_{23}$ are arbitrary real coefficients, and where both the real and imaginary components of $S_1$ are varied as in Table \ref{tab:one}.  Realistic compactifications like that found in the scans of \cite{Ibanez:2007rs} can generate larger hierarchies amongst $S_r$ if desired, e.g. $c_{23} \sim \mathcal{O}(10)$.  We also note that, in the simplified formalism of Section \ref{sec:PHENO}, one may be tempted to relate $S_{2,3}$ to $S_1$ via \eqref{eq:massdifmodel} since one apparently is guaranteed to realize the phenomenologically accurate $\Delta m^2_{ij}$, at least at the scale $M(m_s)$.   However, this is redundant, as the bounds in \eqref{eq:parameterspace}, derived in \cite{Brivio:2017dfq,Brivio:2018rzm}, \emph{already} account for renormalization group evolution (RGE), threshold correction effects, and low-energy experimental bounds.  That is, achieving \eqref{eq:parameterspace} upon transforming to the mass-eigenstate basis of $N_p$ already ensures (within the rough set of assumptions in this analysis and those of  \cite{Brivio:2017dfq,Brivio:2018rzm}) that IR SM and neutrino phenomenology is realized.  Furthermore, the fastidious reader will notice that enforcing the exponential relationship in \eqref{eq:massdifmodel} on $S_{2,3}$ can yield an $M_p$ spectrum that is largely insensitive to orders-of-magnitude changes in $S_1$, and that tuning at the third or fourth digit (or worse) between $S_r$ can be required for $\mathbb{s} \lesssim 13$. Both of these undesirable effects are unnecessary artifacts of the simplified formalism of Section \ref{sec:PHENO}, which while helpful in making our qualitative point analytically, we now relax.  As a final comment, we have confirmed for ourselves in \cite{BTTpert} that the RGE and threshold corrections to $M_p$ down to (e.g.) $\mu \sim m_Z$ are extremely small in comparison to the $\sim$ PeV scales we aim to extract below \cite{Casas:1999tp,Casas:1999tg,Antusch:2005gp,Chankowski:2001mx,Antusch:2003kp,Ibarra:2018dib,Antusch:2002ek,Ibarra:2020eia}.

\begin{table}[t]
\renewcommand{\arraystretch}{1.5}
\noindent
\centering
\begin{tabular}{|c||c|c|c|c|c|c|c|}
\hline
 & $\mathbb{d}_0$ & $\rho$ & $\phi$ & $c_{23}$ & $S_1$ & $\mathbb{s}$  & $N$ \\
\hline 
\hline
\text{Range A} & $-\left[ 5, 3 \right]$ & $\left[ 0,1 \right]$ & $\left[ 0, 2\pi \right]$ & $\left[\frac{1}{10},\frac{15}{10}\right]$ & $( 0, 10\, ]$ &  $ 3\mathbb{Z} \le 18$ & $4$\\
\hline
\text{Range B} & $\left[ \frac{15}{10}, 4 \right]$ & $\left[ 0,1 \right]$ & $\left[ 0, 2\pi \right]$ &  $\left[\frac{1}{10},\frac{15}{10}\right]$ & $( 0, 10\, ]$ &  $3$ & $5$\\
\hline
\text{Range C} & $-\left[ 6, 3 \right]$ & $\left[ 0,1 \right]$ & $\left[ 0, 2\pi \right]$ &  $\left[\frac{1}{10},\frac{15}{10}\right]$ & $(0, 10\, ]$ &  $18$ & $5$\\
\hline
\end{tabular}\, 
\caption{The scan ranges implemented for the relevant free parameters in our formalism.  
}.
\label{tab:one}
\end{table}

\begin{figure}[t]
\includegraphics[width=90mm]{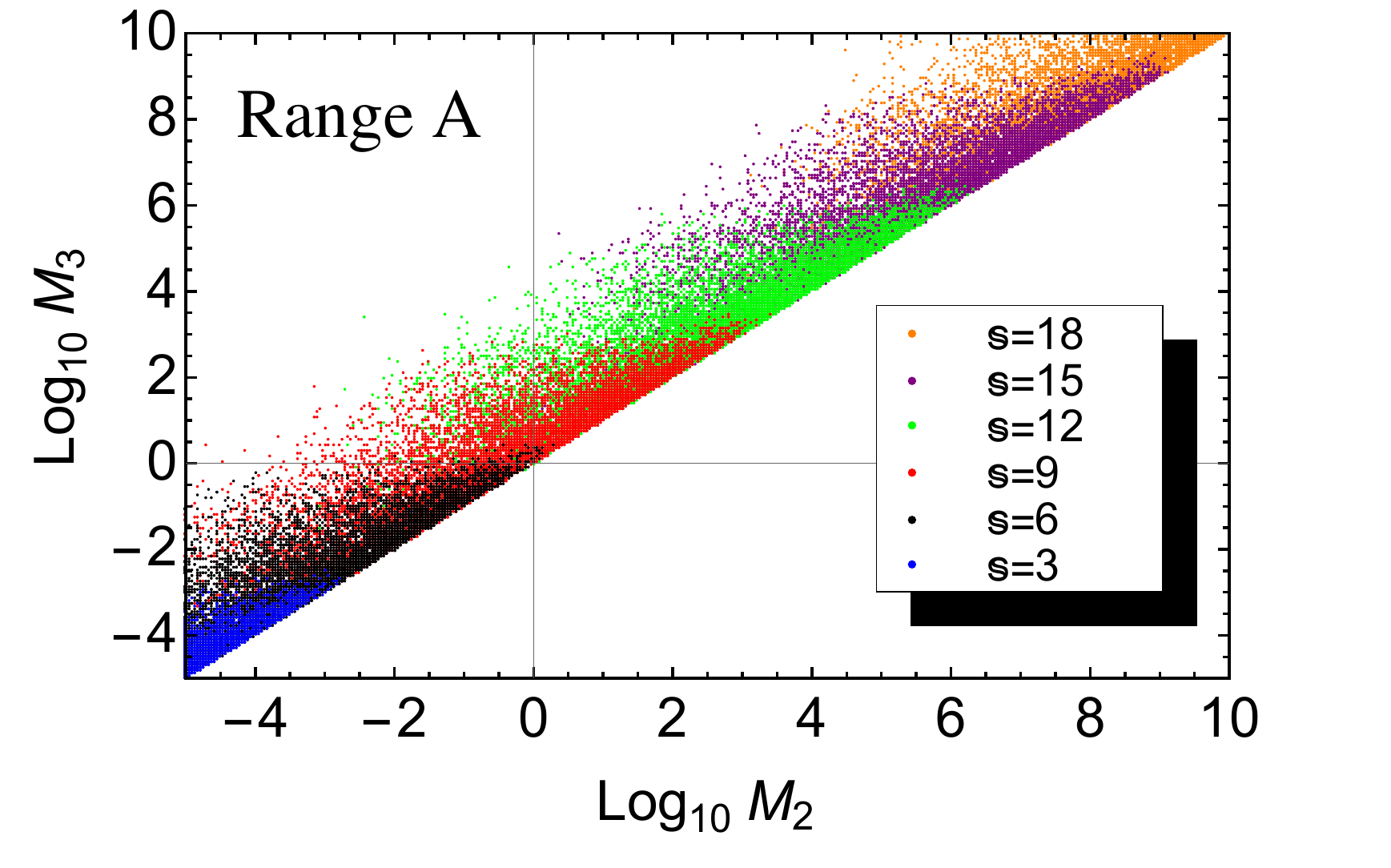}
\includegraphics[width=90mm]{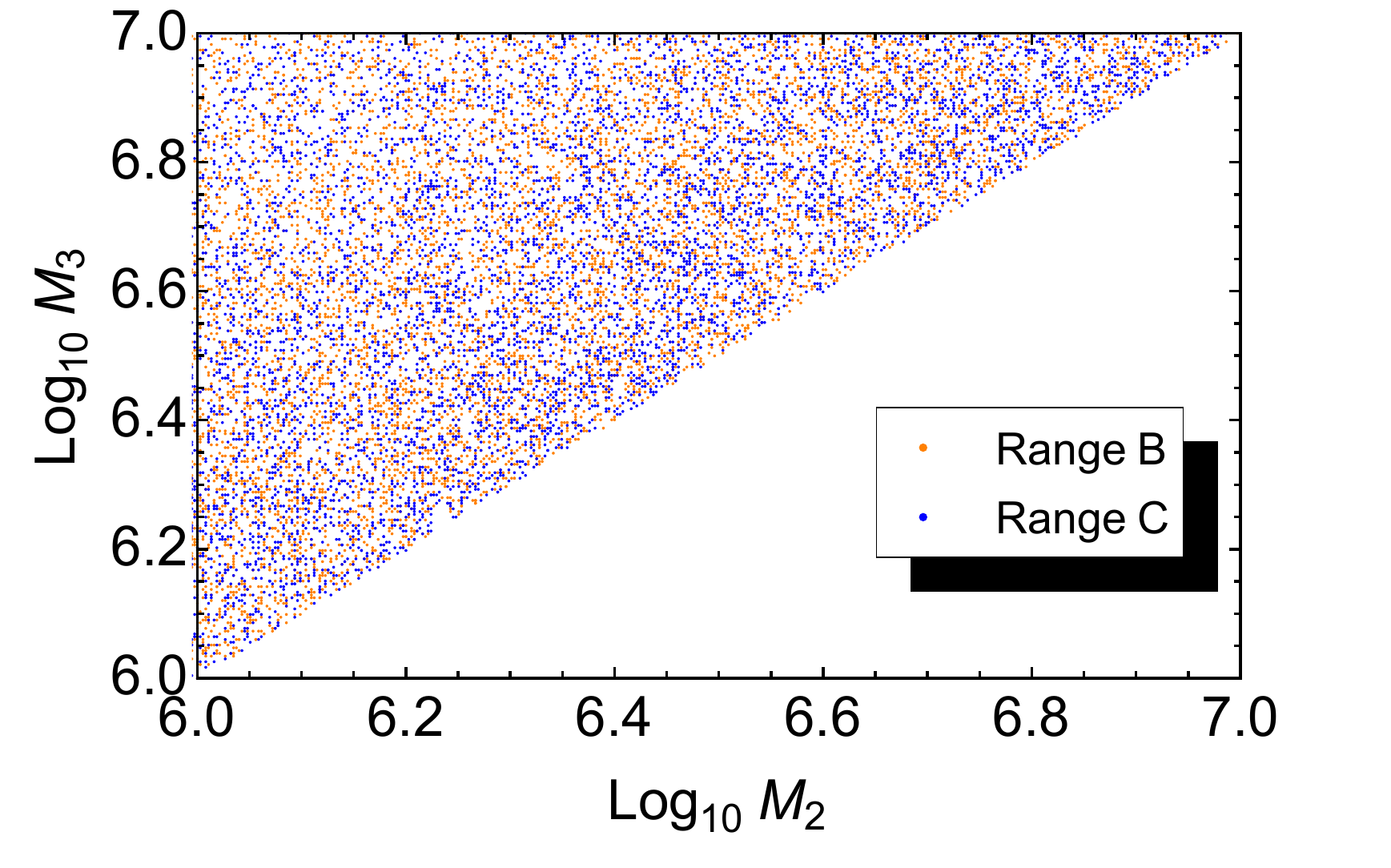}
\caption{TOP:  The spectrum of $\vert M_{2,3} \vert$ given scan Range A from Table \ref{tab:one}.  BOTTOM:  The same for scan Ranges B-C, and in the domain of successful Neutrino Options.}
\label{fig:23spec}
\end{figure}

Given \eqref{eq:dparam}-\eqref{eq:Srelate}, we then perform scans by allowing random variations of $\mathbb{d}_0$, $\rho$, $\phi$, $c_{23}$ and $S_1$ within the ranges identified in Table \ref{tab:one}.  Specifically, our scripts first assemble $M$ with \eqref{eq:MRdem}, given a string scale $m_s$ that we set by hand,
and then we randomly draw the other parameters within their allowed ranges, calculating the magnitudes of the associated $M_p$ spectrum 
for each draw.  We repeat this procedure $10^N$ times, collect all of the inputs, and analyze the outputs.

In the upper plot of Figure \ref{fig:23spec} we show the spectrum of the heaviest $\vert M_{2,3} \vert$ for $\mathbb{s} \in \lbrace 3-18 \rbrace$ given the Range A values of Table \ref{tab:one}.  In this case we use scalings of the flavour vectors motivated by \eqref{eq:djScale}, $-5 \le \mathbb{d}_0 \le -3$, and we note that if these are to truly scale like the neutrino Yukawa couplings $\omega^{(\nu)}$, then intermediate to high values of the string scale, $m_s \sim 10^{12-18}$ GeV, are preferred given \eqref{eq:parameterspace}.  Of course, this conclusion can be changed given a different range of suppression factors $S_r$, although we recall that as discussed in \cite{Ibanez:2006da}, one does not necessarily expect the D-brane instanton actions to be highly suppressed.  Keeping $S_r$ fixed, in the bottom plot of Figure \ref{fig:23spec} we study the more tailored values of the flavour vectors in Range B-C from Table \ref{tab:one}, for the extreme limits of the string scales we consider: $\mathbb{s} = \lbrace 3, 18\rbrace$.  Here we isolate the mass domain to the values preferred in \eqref{eq:parameterspace}, and observe that with this particular subset of $\mathbb{d}_0$, both large and small string-scale scenarios can yield successful Neutrino Options. In all of these models we note that, by construction, the lightest eigenvalue $M_1$ is also of a scale which would maintain natural threshold corrections to the Higgs mass from \eqref{eq:Higgsthresh}. 

Ultimately Figures \ref{fig:contourscale}-\ref{fig:23spec} demonstrate that, in the absence of an explicit string-derived model, or any other further theoretical constraints on the input parameters, the implications of \eqref{eq:MRdem} are not terribly predictive --- a range of eigenvalues spanning most conceivable scales is achievable, including very low values of $M_p$ relevant to, e.g., sterile neutrino dark matter studies.  On the other hand, they also demonstrate that successful, UV-complete Neutrino Options can be achieved in a variety of string embeddings, and further motivate theoretical studies to that end. 
\section{Summary and Outlook}
\label{sec:CONCLUDE}
We have shown that a sterile Majorana mass matrix with eigenvalues $M_p \sim$ PeV can arise naturally through non-perturbative instanton effects, which are known to be present in a large class of string compactifications, including Type IIA non-SUSY intersecting D-brane models. Hence successful Neutrino Options can potentially be realized across a broad range of string scales in this framework, thereby simultaneously explaining the origins of the EW and light neutrino mass scales in a UV-complete way, providing an elegant resolution to (at least) two of the SM's most pressing phenomenological issues.  Furthermore, this mechanism does so without relying on 4D SUSY, although its potential absence puts an important constraint on the types of finite string embeddings that give a consistent framework, free of destabilizing string excitations of SM states. Hence we believe that this mechanism provides an interesting new approach to these important issues in BSM physics and motivates further theoretical exploration of the types of finite string compactifications that can yield \eqref{eq:massgenB}-\eqref{eq:MRdem} and an otherwise stable Higgs mass, any implied constraints on the stringy free parameters at hand (i.e. $S_r$, $d_a$, and $m_s$), and eventually a more detailed phenomenological survey of the resulting neutrino mass and mixing spectrum.
\section*{Acknowledgements}
I gratefully acknowledge support from the Villum Fund, project number 00010102, and thank Stefan Antusch, Luis Ib\'a\~nez, Graham Ross, and Angel Uranga for helpful comments and insight.  I also thank Ilaria Brivio and Michael Trott for inspiring discussions and collaboration on related work, and both Ilaria and Graham for their review of the manuscript.  I also appreciate Ilaria allowing me to recycle the Feynman diagrams in Fig. \ref{fig:match}.


\end{document}